# Study on a Spinorial Representation of Linear Canonical Transformations

**Raoelina Andriambololona[1], Ravo Tokiniaina Ranaivoson[2], Hanitriarivo Rakotoson[3]**

*raoelina.andriambololona@gmail.com[1];raoelinasp@yahoo.fr[1];jacquelineraoelina@hotmail.com[1]; tokhiniaina@gmail.com[2]; infotsara@gmail.com[3]*

*IT and Theoretical Physics Department*

*Institut National des Sciences et Techniques Nucléaires (INSTN- Madagascar)*

BP 3907 Antananarivo 101, MADAGASCAR, *instn@moov.mg*

***Abstract***: This work is a continuation of our previous works concerning linear canonical transformations and phase space representation of quantum theory. It is mainly focused on the description of an approach which allows to establish spinorial representation of linear canonical transformations. This description is started with the presentation of a suitable parameterization of linear canonical transformations which permits to represent them with special pseudo-orthogonal transformations in an operator space. Then the establishment of the spinorial representation is deduced using the well-known relation existing between special pseudo-orthogonal and spin groups. The cases of one dimension and general multidimensional theory are both studied.

## 1. Introduction

In our previous papers [1-4], we have performed a series of study on a phase space representation of quantum theory and Linear Canonical Transformations (LCTs). LCTs have already been studied in various contexts [5-9] but our work is focused on their study in the framework of quantum theory. In the paper [2], we have established that there is an isomorphism between the dispersion operator algebra and the Lie algebra $\mathfrak{sp}(2N_+, 2N_-)$ of the Lie group $Sp(2N_+, 2N_-)$ corresponding to the set of LCTs in a pseudo-euclidian vectorial space of signature $(N_+, N_-)$. This isomorphism permit to build an unitary representation of LCTs. In this paper, our main goal is to describe a method which permits to establish a spinorial representation of LCTs. The approach consists mainly of associating a pseudo-orthogonal transformation to a LCT. Then, the relationships between pseudo-orthogonal group, Clifford algebra and spin groups [10-12] allows the construction of the spinorial representation of LCTs.

In the work [2], we have introduced operators defined from the momentum $\boldsymbol{p}$ and coordinate operators $\boldsymbol{x}$ of a particle. Some of these operators will be used throughout the present paper. These operators are the reduced operators $\boldsymbol{\mathcal{p}}$ and $\boldsymbol{\mathcal{x}}$, the reduced dispersions operators $\beth^{+}, \beth^{-}$ and $\beth^{\times}$ and their multidimensional generalization $\boldsymbol{\mathcal{p}}_{\mu}, \boldsymbol{\mathcal{x}}_{\mu}, \beth_{\mu\nu}^{+}, \beth_{\mu\nu}^{-}$ and $\beth_{\mu\nu}^{\times}$. The notations that we use throughout this paper are inspired from [13].

## 2. Special pseudo-orthogonal transformation associated to a LCT

### 2.1 Case of one dimension theory

In the framework of one dimension quantum mechanics, a LCT is a linear transformation of the form

$$\begin{cases} \boldsymbol{\mathcal{p}}' = \Pi \boldsymbol{\mathcal{p}} + \Theta \boldsymbol{\mathcal{x}} \\ \boldsymbol{\mathcal{x}}' = \Xi \boldsymbol{\mathcal{p}} + \Lambda \boldsymbol{\mathcal{x}} \end{cases} \Leftrightarrow (\boldsymbol{\mathcal{p}}' \quad \boldsymbol{\mathcal{x}}') = (\boldsymbol{\mathcal{p}} \quad \boldsymbol{\mathcal{x}}) \begin{pmatrix} \Pi & \Xi \\ \Theta & \Lambda \end{pmatrix} \tag{2.1}$$

which leaves invariant the canonical commutation relation: $[\boldsymbol{\mathcal{x}}', \boldsymbol{\mathcal{p}}'] = [\boldsymbol{\mathcal{x}}, \boldsymbol{\mathcal{p}}] = i$ . The consequence of this condition is that the matrix $\mathcal{g} = \begin{pmatrix} \Pi & \Xi \\ \Theta & \Lambda \end{pmatrix}$ which describes the transformation must have a determinant equal to 1. $g$ belongs to the Special Linear group $SL(2)$. The Lie algebra $\mathfrak{sl}$ (2) of the Lie group $SL(2)$ is the set of $2 \times 2$ square matrices with trace equal to zero. We choose the parameterization

$$\mathcal{g} = \begin{pmatrix} \Pi & \Xi \\ \Theta & \Lambda \end{pmatrix} = exp[\begin{pmatrix} \mu & \varphi - \theta \\ \varphi + \theta & -\mu \end{pmatrix}] \tag{2.2}$$

To establish a spinorial representation, we have to find a special pseudo-orthogonal representation corresponding to an LCT. We define the following operators

$$\begin{cases} \boldsymbol{\mathcal{p}}^{+} = \frac{1}{\sqrt{2}}(\sigma^{1} \otimes \boldsymbol{\mathcal{p}} + \sigma^{2} \otimes \boldsymbol{\mathcal{x}}) \\ \boldsymbol{\mathcal{x}}^{-} = \frac{1}{\sqrt{2}}(\sigma^{1} \otimes \boldsymbol{\mathcal{x}} - \sigma^{2} \otimes \boldsymbol{\mathcal{p}}) \\ \boldsymbol{\mathcal{x}}^{+} = \frac{1}{\sqrt{2}}(\sigma^{1} \otimes \boldsymbol{\mathcal{x}} + \sigma^{2} \otimes \boldsymbol{\mathcal{p}}) \\ \boldsymbol{\mathcal{p}}^{-} = \frac{1}{\sqrt{2}}(\sigma^{1} \otimes \boldsymbol{\mathcal{p}} - \sigma^{2} \otimes \boldsymbol{\mathcal{x}}) \end{cases} \tag{2.3}$$

in which $\sigma^1$ and $\sigma^2$ are the generators of the Clifford algebra $\mathfrak{C}(2,0) = \mathfrak{C}(2)$ (for instance the Pauli matrices). From the relations (2.1), (2.2) and (2.3) we deduce that for an infinitesimal LCT, the laws of transformations of $\boldsymbol{\mathcal{p}}^{+}, \boldsymbol{\mathcal{p}}^{-}, \boldsymbol{\mathcal{x}}^{+}$ and $\boldsymbol{\mathcal{x}}^{-}$ are

$$\begin{cases} \boldsymbol{\mathcal{p}}'^{+} = \boldsymbol{\mathcal{p}}^{+} + \theta \boldsymbol{\mathcal{x}}^{-} + \varphi \boldsymbol{\mathcal{x}}^{+} + \mu \boldsymbol{\mathcal{p}}^{-} \\ \boldsymbol{\mathcal{x}}'^{-} = -\theta \boldsymbol{\mathcal{p}}^{+} + \boldsymbol{\mathcal{x}}^{-} - \mu \boldsymbol{\mathcal{x}}^{+} + \varphi \boldsymbol{\mathcal{p}}^{-} \\ \boldsymbol{\mathcal{x}}'^{+} = \varphi \boldsymbol{\mathcal{p}}^{+} - \mu \boldsymbol{\mathcal{x}}^{-} + \boldsymbol{\mathcal{x}}^{+} - \theta \boldsymbol{\mathcal{p}}^{-} \\ \boldsymbol{\mathcal{p}}'^{-} = \mu \boldsymbol{\mathcal{p}}^{+} + \varphi \boldsymbol{\mathcal{x}}^{-} + \theta \boldsymbol{\mathcal{x}}^{+} + \boldsymbol{\mathcal{p}}^{-} \end{cases} \tag{2.4}$$

The infinitesimal transformation (2.4) can be put in the matricial form

$$(\boldsymbol{\mathcal{p}}'^{+} \quad \boldsymbol{\mathcal{x}}'^{-} \quad \boldsymbol{\mathcal{x}}'^{+} \quad \boldsymbol{\mathcal{p}}'^{-}) = (\boldsymbol{\mathcal{p}}^{+} \quad \boldsymbol{\mathcal{x}}^{-} \quad \boldsymbol{\mathcal{x}}^{+} \quad \boldsymbol{\mathcal{p}}^{-})(I_4 + \mathcal{X}) \tag{2.5}$$

in which $I_4$ is the $4 \times 4$ identity matrix and $\mathcal{X}$ is the $4 \times 4$ matrix

$$\mathcal{X} = \begin{pmatrix} 0 & -\theta & \varphi & \mu \\ \theta & 0 & -\mu & \varphi \\ \varphi & -\mu & 0 & \theta \\ \mu & \varphi & -\theta & 0 \end{pmatrix} \tag{2.6}$$

It is easy to verify that $\mathcal{X}$ belongs to the Lie algebra $\mathfrak{so}(2,2)$ of the Special pseudo-orthogonal group $SO(2,2)$ i.e $exp(\mathcal{X}) \in SO(2,2)$. It follows from the relations (2.1) (2.2) and (2.3) that the special pseudo- orthogonal transformation defined by $exp(\mathcal{X})$ is associated with the LCT $\mathscr{g}$ defined by (2.1) and (2.2). This correspondence defines a representation of the LCT group with special pseudo-orthogonal transformations on the operator space $\mathbb{E} = \{ (\boldsymbol{p}^+ \quad \boldsymbol{x}^- \quad \boldsymbol{x}^+ \quad \boldsymbol{p}^-)\}$.

**2.2 Case of $N$ −dimensional theory**

As in our work [2], we consider the case of a general linear canonical transformation

$$\begin{cases} \boldsymbol{p_\mu}' = \Pi_\mu^\nu \boldsymbol{p_\nu} + \Theta_\mu^\nu \boldsymbol{x_\nu} \\ \boldsymbol{x_\mu}' = \Xi_\mu^\nu \boldsymbol{p_\nu} + \Lambda_\mu^\nu \boldsymbol{x_\nu} \end{cases} \Leftrightarrow (\boldsymbol{p}' \quad \boldsymbol{x}') = (\boldsymbol{p} \quad \boldsymbol{x}) \begin{pmatrix} \Pi & \Xi \\ \Theta & \Lambda \end{pmatrix} \tag{2.7}$$

which leaves invariant the canonical commutation relations

$$\begin{cases} [\boldsymbol{p}'_\mu, \boldsymbol{p}'_\nu]_- = [\boldsymbol{p_\mu}, \boldsymbol{p_\nu}]_- = 0 \\ [\boldsymbol{x}'_\mu, \boldsymbol{x}'_\nu]_- = [\boldsymbol{x_\mu}, \boldsymbol{x_\nu}]_- = 0 \\ [\boldsymbol{p}'_\mu, \boldsymbol{x}'_\nu]_- = [\boldsymbol{p_\mu}, \boldsymbol{x_\nu}]_- = i\eta_{\mu\nu} \end{cases} \tag{2.8}$$

In these expressions, we have $\mu = 0, \dots N-1$ . $\boldsymbol{p}$ and $\boldsymbol{x}$ are the $1 \times N$ row matrices (covectors) with components $\boldsymbol{p_\mu}$ and $\boldsymbol{x_\mu}$. $\Pi, \Xi, \Theta$ and $\Lambda$ are $N \times N$ square matrices and $\eta_{\mu\nu}$ are the covariant components of the bilinear form with a signature $(N_+, N_-)$ with $N_+ + N_- = N$. Following our paper [2], the $2N \times 2N$ matrix $\mathscr{g} = \begin{pmatrix} \Pi & \Xi \\ \Theta & \Lambda \end{pmatrix}$ belongs to the pseudo-symplectic group $Sp\ (2N_+, 2N_-)$. $\mathscr{g}$ can be written in the form

$$\mathscr{g} = \begin{pmatrix} \Pi & \Xi \\ \Theta & \Lambda \end{pmatrix} = exp[\begin{pmatrix} \lambda + \mu & \varphi - \theta \\ \varphi + \theta & \lambda - \mu \end{pmatrix}] \tag{2.9}$$

in which $\theta, \varphi, \mu$ and $\lambda$ are $N \times N$ matrices which satisfy the following relations

$$\begin{cases} \theta^T = \eta\theta\eta \\ \varphi^T = \eta\varphi\eta \\ \mu^T = \eta\mu\eta \\ \lambda^T = -\eta\lambda\eta \text{ and } Tr(\lambda) = 0 \end{cases} \tag{2.10}$$

The relation (2.9) is the multidimensional generalization of parameterization (2.2). To associate a special pseudo-orthogonal transformation to the LCT defined by $\mathscr{g}$ ,as generalization of the relations (2.3), we introduce the operators

$$\begin{cases} \boldsymbol{p}_\mu^+ = \dfrac{1}{\sqrt{2}}\left(\sigma^1 \otimes \boldsymbol{p}_\mu + \sigma^2 \otimes \boldsymbol{x}_\mu\right) \\ \boldsymbol{x}_\mu^- = \dfrac{1}{\sqrt{2}}\left(\sigma^1 \otimes \boldsymbol{x}_\mu - \sigma^2 \otimes \boldsymbol{p}_\mu\right) \\ \boldsymbol{x}_\mu^+ = \dfrac{1}{\sqrt{2}}\left(\sigma^1 \otimes \boldsymbol{x}_\mu + \sigma^2 \otimes \boldsymbol{p}_\mu\right) \\ \boldsymbol{p}_\mu^- = \dfrac{1}{\sqrt{2}}\left(\sigma^1 \otimes \boldsymbol{p}_\mu - \sigma^2 \otimes \boldsymbol{x}_\mu\right) \end{cases} \tag{2.11}$$

If we denote $\boldsymbol{p}^+, \boldsymbol{x}^-, \boldsymbol{x}^+, \boldsymbol{p}^-$ the $1 \times N$ row matrices (covectors) which admit respectively the operators $\boldsymbol{p}_\mu^+, \boldsymbol{x}_\mu^-, \boldsymbol{x}_\mu^+$ and $\boldsymbol{p}_\mu^-$ as components, we can deduce from the relations (2.7), (2.9) and (2.11) that for an infinitesimal LCT, we have

$$\begin{cases} \boldsymbol{p}'^+ = (I_N + \lambda)\boldsymbol{p}^+ + \theta\boldsymbol{x}^- + \varphi\boldsymbol{x}^+ + \mu\boldsymbol{p}^- \\ \boldsymbol{x}'^- = -\theta\boldsymbol{p}^+ + (I_N + \lambda)\boldsymbol{x}^- - \mu\boldsymbol{x}^+ + \varphi\boldsymbol{p}^- \\ \boldsymbol{x}'^+ = \varphi\boldsymbol{p}^+ - \mu\boldsymbol{x}^- + (I_N + \lambda)\boldsymbol{x}^+ - \theta\boldsymbol{p}^- \\ \boldsymbol{p}'^- = \mu\boldsymbol{p}^+ + \varphi\boldsymbol{x}^- + \theta\boldsymbol{x}^+ + (I_N + \lambda)\boldsymbol{p}^- \end{cases} \tag{2.12}$$

in which $I_N$ is the $N \times N$ identity matrix. The infinitesimal transformation (2.12) can be put in the matricial form

$$(\boldsymbol{p}'^+ \quad \boldsymbol{x}'^- \quad \boldsymbol{x}'^+ \quad \boldsymbol{p}'^-) = (\boldsymbol{p}^+ \quad \boldsymbol{x}^- \quad \boldsymbol{x}^+ \quad \boldsymbol{p}^-)(I_{4N} + \mathcal{X}) \tag{2.13}$$

in which $I_{4N}$ is the $4N \times 4N$ identity matrix and $\mathcal{X}$ is the $4N \times 4N$ square matrix

$$\mathcal{X} = \begin{pmatrix} \lambda & -\theta & \varphi & \mu \\ \theta & \lambda & -\mu & \varphi \\ \varphi & -\mu & \lambda & \theta \\ \mu & \varphi & -\theta & \lambda \end{pmatrix} \tag{2.14}$$

The matrix $\mathcal{X}$ belongs to the Lie algebra $\mathfrak{so}(2N, 2N)$ of the Special pseudo-Orthogonal group $SO(2N, 2N)$ i.e $e^{\mathcal{X}} \in SO(2N, 2N)$. The relation (2.11) then leads to a correspondence between the LCT $\mathscr{g}$ defined by (2.7) and (2.9) and the special pseudo- orthogonal transformation defined by $exp(\mathcal{X})$.

## 3. Spinorial representation of a Linear Canonical Transformation

### 3.1 Definition of the spinorial representation

As established in the previous section, any LCT can be associated with a special pseudo-orthogonal transformation. But any special pseudo-orthogonal transformation has a spinorial representation. It follows that we can construct a spinorial representation of an LCT too. To do so, we introduce then the operator

$$\mathbb{p} = (\alpha_+ \otimes \boldsymbol{p}^+ + \beta_+ \otimes \boldsymbol{x}^- + \beta_- \otimes \boldsymbol{x}^+ + \alpha_- \otimes \boldsymbol{p}^-) \tag{3.1}$$

in which the operators $\boldsymbol{p}^{+}, \boldsymbol{x}^{-}, \boldsymbol{x}^{+}$ and $\boldsymbol{p}^{-}$ are the operator defined in the relation (2.3) and $\alpha_{+}, \beta_{+}, \beta_{-}, \alpha_{-}$ are the generators of the Clifford algebra $\mathfrak{C}(2,2)$. They satisfy the relations

$$\begin{cases} (\alpha_{+})^2 = (\beta_{+})^2 = \mathbb{I} \\ (\beta_{-})^2 = (\alpha_{-})^2 = -\mathbb{I} \\ \alpha_{+}\beta_{+} + \beta_{+}\alpha_{+} = 0 \\ \alpha_{+}\beta_{-} + \beta_{-}\alpha_{+} = 0 \\ \alpha_{+}\alpha_{-} + \alpha_{-}\alpha_{+} = 0 \\ \beta_{+}\alpha_{-} + \alpha_{-}\beta_{+} = 0 \\ \beta_{+}\beta_{-} + \beta_{-}\beta_{+} = 0 \\ \beta_{-}\alpha_{-} + \alpha_{-}\beta_{-} = 0 \end{cases} \quad (3.2)$$

Where $\mathbb{I}$ is the identity in the Clifford algebra $\mathfrak{C}(2.2)$. From the relation (3.2), it can be established that the operators $\alpha_{+}, \beta_{+}, \beta_{-}, \alpha_{-}$ satisfy also the following commutations relations

$$\begin{cases} [\alpha_{+}\beta_{+}, \alpha_{+}]_{-} = -2\beta_{+} \quad [\alpha_{+}\beta_{+}, \beta_{+}]_{-} = 2\alpha_{+} \quad [\alpha_{+}\beta_{+}, \beta_{-}]_{-} = 0 \quad [\alpha_{+}\beta_{+}, \alpha_{-}]_{-} = 0 \\ [\alpha_{+}\beta_{-}, \alpha_{+}]_{-} = -2\beta_{-} \quad [\alpha_{+}\beta_{-}, \beta_{+}]_{-} = 0 \quad [\alpha_{+}\beta_{-}, \beta_{-}]_{-} = -2\alpha_{+} \quad [\alpha_{+}\beta_{-}, \alpha_{-}]_{-} = 0 \\ [\alpha_{+}\alpha_{-}, \alpha_{+}]_{-} = -2\alpha_{-} \quad [\alpha_{+}\alpha_{-}, \beta_{+}]_{-} = 0 \quad [\alpha_{+}\alpha_{-}, \beta_{-}]_{-} = 0 \quad [\alpha_{+}\alpha_{-}, \alpha_{-}]_{-} = -2\alpha_{+} \\ [\beta_{+}\alpha_{-}, \alpha_{+}]_{-} = 0 \quad [\beta_{+}\alpha_{-}, \beta_{+}]_{-} = -2\alpha_{-} \quad [\beta_{+}\alpha_{-}, \beta_{-}]_{-} = 0 \quad [\beta_{+}\alpha_{-}, \alpha_{-}]_{-} = -2\beta_{+} \\ [\beta_{+}\beta_{-}, \alpha_{+}]_{-} = 0 \quad [\beta_{+}\beta_{-}, \beta_{+}]_{-} = -2\beta_{-} \quad [\beta_{+}\beta_{-}, \beta_{-}]_{-} = -2\beta_{+} \quad [\beta_{+}\beta_{-}, \alpha_{-}]_{-} = 0 \\ [\beta_{-}\alpha_{-}, \alpha_{+}]_{-} = 0 \quad [\beta_{-}\alpha_{-}, \beta_{+}]_{-} = 0 \quad [\beta_{-}\alpha_{-}, \beta_{-}]_{-} = 2\alpha_{-} \quad [\beta_{-}\alpha_{-}, \alpha_{-}]_{-} = -2\beta_{-} \end{cases} \quad (3.3)$$

Let us denote $Spin(2,2)\otimes I_2 = \{\mathcal{S} = exp(\vartheta)\otimes I_2\}$ in which $exp(\vartheta)$ is an element of the group $Spin(2,2)$ and $I_2$ the $2 \times 2$ identity matrix. It is easy to remark that the group $Spin(2,2)\otimes I_2$ and $Spin(2,2)$ are isomorph. Let $g = exp\begin{pmatrix} \mu & \varphi - \theta \\ \varphi + \theta & -\mu \end{pmatrix}$ be the element of the LCT group $SL(2)$ corresponding to an LCT (3.1) and $\mathcal{S}$ the element of $Spin(2,2)\otimes I_2$ associated with $g$. This correspondence between $g$ and $\mathcal{S}$ can be described with a mapping $\varrho$ between the LCT group $SL(2)$ and $Spin(2,2)\otimes I_2$ .

$$\mathcal{S} = \varrho(g) \Leftrightarrow \begin{cases} (\boldsymbol{p}' \quad \boldsymbol{x}') = (\boldsymbol{p} \quad \boldsymbol{x})g = (\boldsymbol{p} \quad \boldsymbol{x})exp\begin{pmatrix} \mu & \varphi - \theta \\ \varphi + \theta & -\mu \end{pmatrix} \\ \mathbb{p}' = \mathcal{S}\mathbb{p}\mathcal{S}^{-1} \end{cases} \quad (3.4)$$

$\mathcal{S}$ can be put in the form

$$\begin{aligned} \mathcal{S} &= exp(\vartheta\otimes I_2) \\ &= exp[(\vartheta^1\alpha_{+}\beta_{+} + \vartheta^2\alpha_{+}\beta_{-} + \vartheta^3\alpha_{+}\alpha_{-} + \vartheta^4\beta_{+}\beta_{-} + \vartheta^5\beta_{+}\alpha_{-} + \vartheta^6\beta_{-}\alpha_{-}) \otimes I_2] \end{aligned} \quad (3.5)$$

Using the relation $\mathbb{p}' = \mathcal{S}\mathbb{p}\mathcal{S}^{-1}$ and the relation (3.5), it can be deduced that the law of transformation of $\mathbb{p}$ for an infinitesimal linear canonical transformation is

$$\mathbb{p}' = \mathbb{p} + [\vartheta\otimes I_2, \mathbb{p}]_{-} \quad (3.6)$$

Taking into account the expression (3.11) of $\mathbb{p}$ and the relations (3.3) and (3.6), we deduce the relations

$$\begin{cases} \boldsymbol{p}'^{+} = \boldsymbol{p}^{+} + 2\vartheta^{1}\boldsymbol{x}^{-} - 2\vartheta^{2}\boldsymbol{x}^{+} - 2\vartheta^{3}\boldsymbol{p}^{-} \\ \boldsymbol{x}'^{-} = -2\vartheta^{1}\boldsymbol{p}^{+} + \boldsymbol{x}^{-} - 2\vartheta^{4}\boldsymbol{x}^{+} - 2\vartheta^{5}\boldsymbol{p}^{-} \\ \boldsymbol{x}'^{+} = -2\vartheta^{2}\boldsymbol{p}^{+} - 2\vartheta^{4}\boldsymbol{x}^{-} + \boldsymbol{x}^{+} - 2\vartheta^{6}\boldsymbol{p}^{-} \\ \boldsymbol{p}'^{-} = -2\vartheta^{3}\boldsymbol{p}^{+} - 2\vartheta^{5}\boldsymbol{x}^{-} + 2\vartheta^{6}\boldsymbol{x}^{+} + \boldsymbol{p}^{-} \end{cases} \tag{3.7}$$

The identification of the relation (3.7) with (2.4) permits to deduce the expressions of $\vartheta^1, \vartheta^2, \vartheta^3, \vartheta^4, \vartheta^5$ and $\vartheta^6$ in terms of $\theta, \varphi$ and $\mu$

$$\begin{cases} \vartheta^{1} = \frac{\theta}{2} \quad \vartheta^{2} = -\frac{\varphi}{2} \quad \vartheta^{3} = -\frac{\mu}{2} \\ \vartheta^{4} = \frac{\mu}{2} \quad \vartheta^{5} = -\frac{\varphi}{2} \quad \vartheta^{6} = \frac{\theta}{2} \end{cases} \tag{3.8}$$

The expression (3.5) of $\mathcal{S}$ becomes

$$\begin{aligned} \mathcal{S} &= exp(\vartheta \otimes I_2) \\ &= exp\{[\frac{\theta}{2}(\alpha_+\beta_+ + \beta_-\alpha_-) - \frac{\varphi}{2}(\alpha_+\beta_- + \beta_+\alpha_-) - \frac{\mu}{2}(\alpha_+\alpha_- - \beta_+\beta_-)] \otimes I_2\} \end{aligned} \tag{3.9}$$

The operator $\mathcal{S}$ can be considered as acting on the element $\vec{\psi}$ of a spinor space $\mathbb{S}$

$$\vec{\psi}' = \mathcal{S}(\vec{\psi}) \Leftrightarrow \psi'^{a} = \mathcal{S}_b^a \psi^b \tag{3.10}$$

The couple $(\mathbb{S}, \varrho)$, in which $\varrho$ is the mapping in (3.4), then define a spinorial representation of the LCT group.

### 3.2 Expression of the invariant as polynomial of the operator $\mathbb{p}$

Let us consider the operator $\mathbb{p}$ defined in the relation (3.1) and calculate its square

$$(\mathbb{p})^2 = (\alpha_+ \otimes \boldsymbol{p}^+ + \beta_+ \otimes \boldsymbol{x}^- + \beta_- \otimes \boldsymbol{x}^+ + \alpha_- \otimes \boldsymbol{p}^-)^2 \tag{3.11}$$

From the expression (2.3) of $\boldsymbol{p}^+, \boldsymbol{x}^-, \boldsymbol{x}^+, \boldsymbol{p}^-$, we can deduce the relations

$$\begin{cases} (\boldsymbol{p}^+)^2 + (\boldsymbol{x}^-)^2 - (\boldsymbol{x}^+)^2 - (\boldsymbol{p}^-)^2 = 2\sigma^3 \\ [\boldsymbol{p}^+, \boldsymbol{x}^-]_- = (\boldsymbol{p}^+\boldsymbol{x}^- - \boldsymbol{x}^-\boldsymbol{p}^+) = -4i\beth^+\sigma^3 - i \\ [\boldsymbol{p}^+, \boldsymbol{x}^+]_- = (\boldsymbol{p}^+\boldsymbol{x}^+ - \boldsymbol{x}^+\boldsymbol{p}^+) = 4i\beth^-\sigma^3 \\ [\boldsymbol{p}^+, \boldsymbol{p}^-]_- = (\boldsymbol{p}^+\boldsymbol{p}^- - \boldsymbol{p}^-\boldsymbol{p}^+) = -4i\beth^\times\sigma^3 \\ [\boldsymbol{x}^-, \boldsymbol{x}^+]_- = (\boldsymbol{x}^-\boldsymbol{x}^+ - \boldsymbol{x}^+\boldsymbol{x}^-) = 4i\beth^\times\sigma^3 \\ [\boldsymbol{x}^-, \boldsymbol{p}^-]_- = (\boldsymbol{x}^-\boldsymbol{p}^- - \boldsymbol{p}^-\boldsymbol{x}^-) = 4i\beth^-\sigma^3 \\ [\boldsymbol{x}^+, \boldsymbol{p}^-]_- = (\boldsymbol{p}^-\boldsymbol{x}^+ - \boldsymbol{x}^+\boldsymbol{p}^-) = -4i\beth^+\sigma^3 + i \end{cases} \tag{3.12}$$

in which $\beth^+, \beth^-$ and $\beth^\times$ are the reduced dispersion operators [2]

$$\begin{cases} \beth^+ = \frac{1}{4}[(\boldsymbol{p})^2 + (\boldsymbol{x})^2] \\ \beth^- = \frac{1}{4}[(\boldsymbol{p})^2 - (\boldsymbol{x})^2] \\ \beth^\times = \frac{1}{4}[\boldsymbol{p}\boldsymbol{x} + \boldsymbol{x}\boldsymbol{p}] \end{cases} \tag{3.13}$$

Using the relations (3.2), (3.12) and (3.13) we can deduce the expression of $(\not{p})^2$

$$(\not{p})^2 = \mathbb{D} + 2\mathbb{I} \otimes \sigma^3 - i(\alpha_+\beta_+ - \beta_-\alpha_-) \otimes I_2 \tag{3.14}$$

in which $\mathbb{D}$ is the operator

$$\begin{aligned}\mathbb{D} = -4i[(\alpha_+\beta_+ + \beta_-\alpha_-) \otimes \sigma^3 \otimes \beth^+ - (\alpha_+\beta_- + \beta_+\alpha_-) \otimes \sigma^3 \otimes \beth^-\\ +(\alpha_+\alpha_- - \beta_+\beta_-) \otimes \sigma^3 \otimes \beth^\times]\end{aligned} \tag{3.15}$$

$\mathbb{I}$ is the identity operator in the Clifford algebra $\mathfrak{C}(2.2)$ and $I_2$ is $2 \times 2$ identity matrix. If we introduce the operators

$$\begin{cases}\mho_+ = \frac{1}{2}(\alpha_+\beta_+ + \beta_-\alpha_-)\otimes\sigma^3 \\ \mho_- = \frac{1}{2}(\alpha_+\beta_- + \beta_+\alpha_-)\otimes\sigma^3 \\ \mho_\times = \frac{1}{2}(\alpha_+\alpha_- - \beta_+\beta_-)\otimes\sigma^3 \\ \mho_o = \frac{1}{2}(\alpha_+\beta_+ - \beta_-\alpha_-)\otimes\sigma^3\end{cases} \tag{3.16}$$

we obtain as expression of $\mathbb{D}$

$$\mathbb{D} = -8i[\mho_+\otimes\beth^+ - \mho_-\otimes\beth^- + \mho_\times\otimes\beth^\times] \tag{3.17}$$

using the fact that $(\sigma^3)^2 = I_2$, we may write also for the expression (3.15) of $(\not{p})^2$ and the expression (3.10) of $\mathcal{S}$

$$(\not{p})^2 = \mathbb{D} + 2(\mathbb{I} \otimes \sigma^3)(\mathbb{I} \otimes I_2 - i\mho_o) \tag{3.18}$$

$$\mathcal{S} = exp(\vartheta\otimes I_2) = exp[(\theta\mho_+ - \varphi\mho_- - \mu\mho_\times)(\mathbb{I} \otimes \sigma^3)] \tag{3.19}$$

It can be deduced from the relations (3.2) and (3.16) that the operators $\mho_+, \mho_-, \mho_\times$ and $\mho_o$ satisfy the relation

$$\begin{cases}(\mho_+)^2 = \frac{1}{2}(\epsilon - \mathbb{I}) \otimes I_2 \quad (\mho_-)^2 = -\frac{1}{2}(\epsilon - \mathbb{I}) \otimes I_2 \\ (\mho_\times)^2 = -\frac{1}{2}(\epsilon - \mathbb{I}) \otimes I_2 \quad (\mho_o)^2 = -\frac{1}{2}(\epsilon + \mathbb{I}) \otimes I_2 \\ \mho_+\mho_- = -\mho_-\mho_+ = (\mathbb{I} \otimes \sigma^3)\mho_\times \\ \mho_-\mho_\times = -\mho_\times\mho_- = -(\mathbb{I} \otimes \sigma^3)\mho_+ \\ \mho_\times\mho_+ = -\mho_+\mho_\times = (\mathbb{I} \otimes \sigma^3)\mho_- \\ \mho_+\mho_o = \mho_o\mho_+ = 0 \\ \mho_-\mho_o = \mho_o\mho_- = 0 \\ \mho_+\mho_o = \mho_o\mho_+ = 0\end{cases} \tag{3.20}$$

in which

$$\epsilon = \alpha_+\beta_+\beta_-\alpha_- \tag{3.21}$$

According to the relation (3.5), the law of transformation of $\mathbb{p}$ is $\mathbb{p}' = \mathcal{S}\mathbb{p}\mathcal{S}^{-1}$. It follows that we have for the law of transformation of $(\mathbb{p})^2$

$$(\mathbb{p}')^2 = \mathbb{p}'\mathbb{p}' = \mathcal{S}\mathbb{p}\mathcal{S}^{-1}\mathcal{S}\mathbb{p}\mathcal{S}^{-1} = \mathcal{S}(\mathbb{p})^2\mathcal{S}^{-1}$$
$$= \mathcal{S}\mathbb{D}\mathcal{S}^{-1} + (2\mathbb{I}\otimes\sigma^3)(\mathbb{I}\otimes I_2 - i\mho_o) \qquad (3.22)$$

as we have $\mathcal{S}\mathbb{D}\mathcal{S}^{-1} \neq \mathbb{D}$, it follows that $(\mathbb{p}')^2 \neq (\mathbb{p})^2$ i.e $(\mathbb{p})^2$ is not an invariant. However, it can be shown that an invariant is the polynomial of 4th degree $(\mathbb{p})^4 + 4(\mathbb{I}\otimes\sigma^3)(\mathbb{p})^2$ i.e we have the relation

$$(\mathbb{p}')^4 + 4(\mathbb{I}\otimes\sigma^3)(\mathbb{p}')^2 = \mathcal{S}[(\mathbb{p})^4 + 4(\mathbb{I}\otimes\sigma^3)(\mathbb{p})^2]\mathcal{S}^{-1} = (\mathbb{p})^4 + 4(\mathbb{I}\otimes\sigma^3)(\mathbb{p})^2 \qquad (3.23)$$

### 3.3 Case of multidimensional theory

According to the relations (2.12), (2.13) and (2.14), the special pseudo-orthogonal transformation corresponding to the LCT defined by the relations (2.7), (2.8) and (2.9) is an element of the group $SO(2N,2N)$ so the spin group which is to be used to construct the spinorial representation is $Spin\ (2N,2N)$.Generalizing the relation (3.1), we introduce

$$\mathbb{p} = \alpha_+^\mu \otimes \boldsymbol{p}_\mu^+ + \beta_+^\mu \otimes \boldsymbol{x}_\mu^- + \beta_-^\mu \otimes \boldsymbol{x}_\mu^+ + \alpha_-^\mu \otimes \boldsymbol{p}_\mu^- \qquad (3.24)$$

in which $\boldsymbol{p}_\mu^+, \boldsymbol{x}_\mu^-, \boldsymbol{x}_\mu^+$ and $\boldsymbol{p}_\mu^-$ are the operators defined in the relation (2.11) and the $4N$ operators $\alpha_+^\mu, \beta_+^\mu, \beta_-^\mu, \alpha_-^\mu (\mu = 0, 1, \ldots, N-1)$ are the generators of the Clifford algebra $\mathfrak{C}(2N,2N)$. They are characterized by the relations

$$\begin{cases} \alpha_+^\mu\alpha_+^\nu + \alpha_+^\nu\alpha_+^\mu = 2\eta^{\mu\nu}\mathbb{I} & \beta_+^\mu\beta_+^\nu + \beta_+^\nu\beta_+^\mu = 2\eta^{\mu\nu}\mathbb{I} \\ \beta_-^\mu\beta_-^\nu + \beta_-^\nu\beta_-^\mu = -2\eta^{\mu\nu}\mathbb{I} & \alpha_-^\mu\alpha_-^\nu + \alpha_-^\nu\alpha_-^\mu = -2\eta^{\mu\nu}\mathbb{I} \\ \alpha_+^\mu\beta_+^\nu + \beta_+^\nu\alpha_+^\mu = 0 & \alpha_+^\mu\beta_-^\nu + \beta_-^\nu\alpha_+^\mu = 0 \\ \alpha_+^\mu\alpha_-^\nu + \alpha_-^\nu\alpha_+^\mu = 0 & \beta_+^\mu\beta_-^\nu + \beta_-^\nu\beta_+^\mu = 0 \\ \beta_+^\mu\alpha_-^\nu + \alpha_-^\nu\beta_+^\mu = 0 & \beta_-^\mu\alpha_-^\nu + \alpha_-^\nu\beta_-^\mu = 0 \end{cases} \qquad (3.25)$$

in which $\mathbb{I}$ is the identity of the Clifford algebra $\mathfrak{C}(2N,2N)$. It can be established that the operators $\alpha_+^\mu, \beta_+^\mu, \beta_-^\mu, \alpha_-^\mu$ satisfy the following commutations relations

$$\begin{cases} [\alpha_+^\mu\alpha_+^\nu, \alpha_+^\rho]_- = 2\eta^{\nu\rho}\alpha_+^\mu & [\alpha_+^\mu\alpha_+^\nu, \beta_+^\rho]_- = 0 & [\alpha_+^\mu\alpha_+^\nu, \beta_-^\rho]_- = 0 & [\alpha_+^\mu\alpha_+^\nu, \alpha_-^\rho]_- = 0 \\ [\beta_+^\mu\beta_+^\nu, \alpha_+^\rho]_- = 0 & [\beta_+^\mu\beta_+^\nu, \beta_+^\rho]_- = 2\eta^{\nu\rho}\beta_+^\mu & [\beta_+^\mu\beta_+^\nu, \beta_-^\rho]_- = 0 & [\beta_+^\mu\beta_+^\nu, \alpha_-^\rho]_- = 0 \\ [\beta_-^\mu\beta_-^\nu, \alpha_+^\rho]_- = 0 & [\beta_-^\mu\beta_-^\nu, \beta_+^\rho]_- = 0 & [\beta_-^\mu\beta_-^\nu, \beta_-^\rho]_- = -2\eta^{\nu\rho}\beta_-^\mu & [\beta_-^\mu\beta_-^\nu, \alpha_-^\rho]_- = 0 \\ [\alpha_-^\mu\alpha_-^\nu, \alpha_+^\rho]_- = 0 & [\alpha_-^\mu\alpha_-^\nu, \beta_+^\rho]_- = 0 & [\alpha_-^\mu\alpha_-^\nu, \beta_-^\rho]_- = 0 & [\alpha_-^\mu\alpha_-^\nu, \alpha_-^\rho]_- = -2\eta^{\nu\rho}\alpha_-^\mu \\ [\alpha_+^\mu\alpha_-^\nu, \alpha_+^\rho]_- = -2\eta^{\mu\rho}\alpha_-^\nu & [\alpha_+^\mu\alpha_-^\nu, \beta_+^\rho]_- = 0 & [\alpha_+^\mu\alpha_-^\nu, \beta_-^\rho]_- = 0 & [\alpha_+^\mu\alpha_-^\nu, \alpha_-^\rho]_- = 2\eta^{\nu\rho}\alpha_+^\mu \\ [\alpha_+^\mu\beta_+^\nu, \alpha_+^\rho]_- = -2\eta^{\mu\rho}\beta_+^\nu & [\alpha_+^\mu\beta_+^\nu, \beta_+^\rho]_- = 2\eta^{\nu\rho}\alpha_+^\mu & [\alpha_+^\mu\beta_+^\nu, \beta_-^\rho]_- = 0 & [\alpha_+^\mu\beta_+^\nu, \alpha_-^\rho]_- = 0 \\ [\alpha_+^\mu\beta_-^\nu, \alpha_+^\rho]_- = -2\eta^{\mu\rho}\beta_-^\nu & [\alpha_+^\mu\beta_-^\nu, \beta_+^\rho]_- = 0 & [\alpha_+^\mu\beta_-^\nu, \beta_-^\rho]_- = -2\eta^{\nu\rho}\alpha_+^\mu & [\alpha_+^\mu\beta_-^\nu, \alpha_-^\rho]_- = 0 \\ [\beta_+^\mu\beta_-^\nu, \alpha_+^\rho]_- = 0 & [\beta_+^\mu\beta_-^\nu, \beta_+^\rho]_- = -2\eta^{\mu\rho}\beta_+^\nu & [\beta_+^\mu\beta_-^\nu, \beta_-^\rho]_- = -2\eta^{\nu\rho}\beta_+^\mu & [\beta_+^\mu\beta_-^\nu, \alpha_-^\rho]_- = 0 \\ [\beta_+^\mu\alpha_-^\nu, \alpha_+^\rho]_- = 0 & [\beta_+^\mu\alpha_-^\nu, \beta_+^\rho]_- = 2\eta^{\mu\rho}\alpha_-^\nu & [\beta_+^\mu\alpha_-^\nu, \beta_-^\rho]_- = 0 & [\beta_+^\mu\alpha_-^\nu, \alpha_-^\rho]_- = -2\eta^{\nu\rho}\beta_+^\mu \\ [\beta_-^\mu\alpha_-^\nu, \alpha_+^\rho]_- = 0 & [\beta_-^\mu\alpha_-^\nu, \alpha_-^\rho]_- = -2\eta^{\nu\rho}\beta_+^\mu & [\beta_-^\mu\alpha_-^\nu, \beta_-^\rho]_- = 2\eta^{\mu\rho}\alpha_-^\nu & [\beta_-^\mu\alpha_-^\nu, \beta_-^\rho]_- = 0 \end{cases} \qquad (3.26)$$

Let us denote $Spin(2N,2N)\otimes I_2=\{\mathcal{S}=exp(\vartheta)\otimes I_2=exp(\vartheta\otimes I_2)\}$ in which $exp(\vartheta)$ is an element of $Spin(2N,2N)$ and $I_2$the $2\times 2$ identity matrix. Let $\mathcal{g}=exp[\begin{pmatrix}\lambda+\mu & \varphi-\theta\\ \varphi+\theta & \lambda-\mu\end{pmatrix}]$ be the element of the LCT group $Sp(2N_+,2N_-)$ corresponding to an LCT (3.7) and $\mathcal{S}$ the element of $Spin(2N,2N)\otimes I_2$ associated with $\mathcal{g}$. The correspondence between $g$ and $\mathcal{S}$ can be described with a mapping $\varrho$ between the LCT group $Sp(2N_+,2N_-)$ and $Spin(2N,2N)\otimes I_2$.

$$\mathcal{S}=\varrho(\mathcal{g})\Leftrightarrow\begin{cases}(\boldsymbol{p}' \quad \boldsymbol{x}')=(\boldsymbol{p} \quad \boldsymbol{x})\mathcal{g}=(\boldsymbol{p} \quad \boldsymbol{x})exp[\begin{pmatrix}\lambda+\mu & \varphi-\theta\\ \varphi+\theta & \lambda-\mu\end{pmatrix}]\\ \mathbb{p}'=\mathcal{S}\mathbb{p}\mathcal{S}^{-1}\end{cases} \tag{3.27}$$

For an infinitesimal LCT, the law of transformation of $\mathbb{p}$ is

$$\mathbb{p}'=\mathbb{p}+[\vartheta\otimes I_2,\mathbb{p}]_- \tag{3.28}$$

Using the fact that $\vartheta$ belongs to the Lie algebra of $Spin(2N,2N)$ and taking into account the relations (2.11), (2.12) (3.24), (3.26) and (3.28) we can find the expression of $\vartheta$

$$\begin{aligned}\vartheta=\eta_{\nu\rho}[&\frac{\theta_\mu^\rho}{2}(\alpha_+^\mu\beta_+^\nu+\beta_-^\nu\alpha_-^\mu)-\frac{\varphi_\mu^\rho}{2}(\alpha_+^\mu\beta_-^\nu+\beta_+^\nu\alpha_-^\mu)+\frac{\mu_\mu^\rho}{2}(\alpha_+^\mu\alpha_-^\nu+\beta_+^\nu\beta_-^\mu)\\ &+\frac{\lambda_\mu^\rho}{2}\left(\alpha_+^\mu\alpha_+^\nu+\beta_+^\mu\beta_+^\nu-\beta_-^\mu\beta_-^\nu-\alpha_-^\mu\alpha_-^\nu\right)]\end{aligned} \tag{3.29}$$

then we have for $\mathcal{S}$

$$\begin{aligned}\mathcal{S}=exp[\eta_{\nu\rho}[&\frac{\theta_\mu^\rho}{2}(\alpha_+^\mu\beta_+^\nu+\beta_-^\nu\alpha_-^\mu)-\frac{\varphi_\mu^\rho}{2}(\alpha_+^\mu\beta_-^\nu+\beta_+^\nu\alpha_-^\mu)+\frac{\mu_\mu^\rho}{2}(\alpha_+^\mu\alpha_-^\nu+\beta_+^\nu\beta_-^\mu)\\ &+\frac{\lambda_\mu^\rho}{2}\left(\alpha_+^\mu\alpha_+^\nu+\beta_+^\mu\beta_+^\nu-\beta_-^\mu\beta_-^\nu-\alpha_-^\mu\alpha_-^\nu\right)]\otimes I_2]\end{aligned} \tag{3.30}$$

The operator $\mathcal{S}$ can be considered as acting on the spinors $\vec{\psi}$ of a spinor space $\mathbb{S}$

$$\vec{\psi}'=\mathcal{S}\vec{\psi}\Leftrightarrow\psi'^a=\mathcal{S}_b^a\psi^b \tag{3.31}$$

The couple $(\mathbb{S},\varrho)$ define the spinorial representation of the LCT group.
Using the relation (3.11), it can be established that the operators $\boldsymbol{p}_\mu^+,\boldsymbol{x}_\mu^-,\boldsymbol{x}_\mu^+$ and $\boldsymbol{p}_\mu^-$ satisfy the relations

$$\begin{cases}\boldsymbol{p}_\mu^+\boldsymbol{p}_\nu^+=2\beth_{\boldsymbol{\mu\nu}}^++i\sigma^3\left(\beth_{\boldsymbol{\mu\nu}}^\times-\beth_{\boldsymbol{\nu\mu}}^\times\right)-\frac{1}{2}\sigma^3\eta_{\mu\nu}=2\beth_{\boldsymbol{\mu\nu}}^++2i\sigma^3\beth_{\boldsymbol{\mu\nu}}^\ltimes-\frac{1}{2}\sigma^3\eta_{\mu\nu}\\ \boldsymbol{x}_\mu^-\boldsymbol{x}_\nu^-=2\beth_{\boldsymbol{\mu\nu}}^++i\sigma^3(\beth_{\boldsymbol{\mu\nu}}^\times-\beth_{\boldsymbol{\nu\mu}}^\times)-\frac{1}{2}\sigma^3\eta_{\mu\nu}=2\beth_{\boldsymbol{\mu\nu}}^++2i\sigma^3\beth_{\boldsymbol{\mu\nu}}^\ltimes-\frac{1}{2}\sigma^3\eta_{\mu\nu}\\ \boldsymbol{x}_\mu^+\boldsymbol{x}_\nu^+=2\beth_{\boldsymbol{\mu\nu}}^+-i\sigma^3\left(\beth_{\boldsymbol{\mu\nu}}^\times-\beth_{\boldsymbol{\nu\mu}}^\times\right)+\frac{1}{2}\sigma^3\eta_{\mu\nu}=2\beth_{\boldsymbol{\mu\nu}}^+-2i\sigma^3\beth_{\boldsymbol{\mu\nu}}^\ltimes+\frac{1}{2}\sigma^3\eta_{\mu\nu}\\ \boldsymbol{p}_\mu^-\boldsymbol{p}_\nu^-=2\beth_{\boldsymbol{\mu\nu}}^+-i\sigma^3\left(\beth_{\boldsymbol{\mu\nu}}^\times-\beth_{\boldsymbol{\nu\mu}}^\times\right)+\frac{1}{2}\sigma^3\eta_{\mu\nu}=2\beth_{\boldsymbol{\mu\nu}}^+-2i\sigma^3\beth_{\boldsymbol{\mu\nu}}^\ltimes+\frac{1}{2}\sigma^3\eta_{\mu\nu}\\ [\boldsymbol{p}_\mu^+,\boldsymbol{x}_\nu^-]_-=\boldsymbol{p}_\mu^+\boldsymbol{x}_\nu^--\boldsymbol{x}_\nu^-\boldsymbol{p}_\mu^+=-4i\sigma^3\beth_{\mu\nu}^++i\eta_{\boldsymbol{\mu\nu}}\\ [\boldsymbol{p}_\mu^+,\boldsymbol{x}_\nu^+]_-=\boldsymbol{p}_\mu^+\boldsymbol{x}_\nu^+-\boldsymbol{x}_\nu^+\boldsymbol{p}_\mu^+=4i\sigma^3\beth_{\mu\nu}^-\\ [\boldsymbol{p}_\mu^+,\boldsymbol{p}_\nu^-]_-=\boldsymbol{p}_\mu^+\boldsymbol{p}_\nu^--\boldsymbol{p}_\nu^-\boldsymbol{p}_\mu^+=-2i\sigma^3(\beth_{\boldsymbol{\mu\nu}}^\times+\beth_{\boldsymbol{\nu\mu}}^\times)=-4i\sigma^3\beth_{\mu\nu}^\bowtie\\ [\boldsymbol{x}_\nu^-,\boldsymbol{x}_\mu^+]_-=\boldsymbol{x}_\nu^-\boldsymbol{x}_\mu^+-\boldsymbol{x}_\mu^+\boldsymbol{x}_\nu^-=2i\sigma^3(\beth_{\mu\nu}^\times+\beth_{\nu\mu}^\times)=4i\sigma^3\beth_{\mu\nu}^\bowtie\\ [\boldsymbol{x}_\nu^-,\boldsymbol{p}_\mu^-]_-=\boldsymbol{x}_\nu^-\boldsymbol{p}_\mu^--\boldsymbol{p}_\mu^-\boldsymbol{x}_\nu^-=-4i\sigma^3\beth_{\mu\nu}^-\\ [\boldsymbol{x}_\nu^+,\boldsymbol{p}_\mu^-]_-=\boldsymbol{x}_\nu^+\boldsymbol{p}_\mu^--\boldsymbol{p}_\mu^-\boldsymbol{x}_\nu^+=-4i\sigma^3\beth_{\mu\nu}^+-i\eta_{\boldsymbol{\mu\nu}}\end{cases} \tag{3.32}$$

in which [2]

$$
\begin{cases}
\daleth_{\mu\nu}^{+} = \dfrac{1}{4}(\boldsymbol{p}_{\mu}\boldsymbol{p}_{\nu} + \boldsymbol{x}_{\mu}\boldsymbol{x}_{\nu}) \\
\daleth_{\mu\nu}^{-} = \dfrac{1}{4}(\boldsymbol{p}_{\mu}\boldsymbol{p}_{\nu} - \boldsymbol{x}_{\mu}\boldsymbol{x}_{\nu}) \\
\daleth_{\mu\nu}^{\times} = \dfrac{1}{4}\left(\boldsymbol{p}_{\mu}\boldsymbol{x}_{\nu} + \boldsymbol{x}_{\nu}\boldsymbol{p}_{\mu}\right) = \daleth_{\mu\nu}^{\bowtie} + \daleth_{\boldsymbol{\mu\nu}}^{\ltimes} \\
\daleth_{\mu\nu}^{\bowtie} = \dfrac{1}{2}\left(\daleth_{\mu\nu}^{\times} + \daleth_{\nu\mu}^{\times}\right) = \dfrac{1}{8}\left(\boldsymbol{p}_{\mu}\boldsymbol{x}_{\nu} + \boldsymbol{x}_{\nu}\boldsymbol{p}_{\mu} + \boldsymbol{p}_{\nu}\boldsymbol{x}_{\mu} + \boldsymbol{x}_{\mu}\boldsymbol{p}_{\nu}\right) \\
\daleth_{\boldsymbol{\mu\nu}}^{\ltimes} = \dfrac{1}{2}\left(\daleth_{\mu\nu}^{\times} - \daleth_{\nu\mu}^{\times}\right) = \dfrac{1}{8}\left(\boldsymbol{p}_{\mu}\boldsymbol{x}_{\nu} + \boldsymbol{x}_{\nu}\boldsymbol{p}_{\mu} - \boldsymbol{p}_{\nu}\boldsymbol{x}_{\mu} - \boldsymbol{x}_{\mu}\boldsymbol{p}_{\nu}\right)
\end{cases}
\tag{3.33}
$$

From the relations (3.31) and (3.32), we have for the square$(\mathbb{p})^2$ of the operator $\mathbb{p}$

$$
\begin{aligned}
(\mathbb{p})^2 &= (\alpha_{+}^{\mu} \otimes \boldsymbol{p}_{\mu}^{+} + \beta_{+}^{\mu} \otimes \boldsymbol{x}_{\mu}^{-} + \beta_{-}^{\mu} \otimes \boldsymbol{x}_{\mu}^{+} + \alpha_{-}^{\mu} \otimes \boldsymbol{p}_{\mu}^{-} )^2 \\
&= \mathbb{D} - 2(\mathbb{I} \otimes \sigma^3)(N\mathbb{I} \otimes I_2 - i\eta_{\mu\nu}\mho_{o}^{\mu\nu})
\end{aligned}
\tag{3.34}
$$

in which

$$
\mathbb{D} = -8i[\mho_{+}^{\mu\nu} \otimes \daleth_{\mu\nu}^{+} - \mho_{-}^{\mu\nu} \otimes \daleth_{\mu\nu}^{-} + \mho_{\bowtie}^{\mu\nu} \otimes \daleth_{\mu\nu}^{\bowtie} - \daleth_{\boldsymbol{\mu\nu}}^{\ltimes} \otimes \mho_{\ltimes}^{\mu\nu}]
\tag{3.35}
$$

$$
\begin{cases}
\mho_{+}^{\mu\nu} = \dfrac{1}{2}[\left(\alpha_{+}^{\mu}\beta_{+}^{\nu} + \beta_{-}^{\nu}\alpha_{-}^{\mu}\right) \otimes \sigma^3] \\
\mho_{-}^{\mu\nu} = \dfrac{1}{2}[\left(\alpha_{+}^{\mu}\beta_{-}^{\nu} + \beta_{+}^{\nu}\alpha_{-}^{\mu}\right) \otimes \sigma^3] \\
\mho_{\bowtie}^{\mu\nu} = \dfrac{1}{2}[\left(\alpha_{+}^{\mu}\alpha_{-}^{\nu} + \beta_{+}^{\nu}\beta_{-}^{\mu}\right) \otimes \sigma^3] \\
\mho_{\ltimes}^{\mu\nu} = \dfrac{1}{4}\left[(\alpha_{+}^{\mu}\alpha_{+}^{\nu} + \beta_{+}^{\mu}\beta_{+}^{\nu} - \beta_{-}^{\mu}\beta_{-}^{\nu} - \alpha_{-}^{\mu}\alpha_{-}^{\nu}) \otimes \sigma^3\right] \\
\mho_{o}^{\mu\nu} = \dfrac{1}{2}[(\alpha_{+}^{\mu}\beta_{+}^{\nu} - \beta_{-}^{\nu}\alpha_{-}^{\mu}) \otimes \sigma^3]
\end{cases}
\tag{3.36}
$$

Using the relation (3.36), the expression of the operator $\mathcal{S}$ in (3.30) may be written in a compact form

$$
\mathcal{S} = exp\left[\eta_{\nu\rho}\left(\theta_{\mu}^{\rho}\mho_{+}^{\mu\nu} - \varphi_{\mu}^{\rho}\mho_{+}^{\mu\nu} + \mu_{\mu}^{\rho}\mho_{\bowtie}^{\mu\nu} + \lambda_{\mu}^{\rho}\mho_{\ltimes}^{\mu\nu}\right)(\mathbb{I} \otimes \sigma^3)\right]
\tag{3.37}
$$

The operator $\mathbb{D}$ defined by the relation (3.35) generalizes the one defined, for one dimension, in (3.15),(3.17) . It is obvious to note that the operators$(\mathbb{p})^2$ and $\mathbb{D}$ are not invariant. The case of one dimension theory, studied in the section 3.2, suggests that an invariant may be polynomial function in $\mathbb{p}$ with a degree greater than or equal to 4.

## 4-Conclusion

The approach described in this work shows that it is possible to establish a spinorial representation of Linear Canonical Transformations. As it is shown in the sections 2 and 3, the establishment of this spinorial representation follows from the fact that, using an adequate parameterization, it is possible to associate with a linear canonical transformation a special pseudo-orthogonal transformation in an operator space. Then the spinorial representation can be established easily using the relationship between special pseudo-orthogonal group and spin group.

A main result thus obtained is the explicit expression of the operator $\mathcal{S}$ which corresponds to the representation of an LCT in a spinor space. It is given in the relation (3.9) or (3.19) for the case of one dimension theory and in the relation (3.30) or (3.37) for the multidimensional case.

Our study leads to the introduction of the operator $\not{\mathbb{p}}$ which is given respectively for one dimension theory and multidimensional cases in the relations (3.1), and (3.24). As shown by the relations (3.4), (3.27), this operator is useful to write the explicit expression of the transformation corresponding to the spinorial representation of the LCT. According to the relation (3.23), the invariant is not $(\not{\mathbb{p}})^2$, as it may be expected for a spinorial representation on an ordinary commutative vector space, but a higher degree polynomial function in $\not{\mathbb{p}}$. This result is probably a consequence of the fact that the space on which the LCT is defined is a noncommutative operator space.

According to our works [14] and [15], LCTs can be linked with many interesting physical problems like the generalization of Fourier and Lorentz transformations in the framework of a relativistic quantum theory and the study of the properties of elementary fermions of the Standard Model. Following this way, we have, for instance, established in [15] a new approach to explain the charges (electric charge, hypercharge and colors) of the elementary fermions. It was shown that the electric charge can be written as the sum of four terms, the weak hypercharge of five terms and the weak isospin of two terms. These facts suggest us to expect that the results established in the present paper may have many interesting application in quantum theory and related domains.